\documentclass{revtex4}

\usepackage{color}
\usepackage{graphicx}
\usepackage{dcolumn}
\usepackage{amsmath}

\begin{document}
\bibliographystyle{revtex}


\title[Short Title]{Isospin-asymmetric nuclear matter}

\author{J.A. L\'opez, E. Ram\'irez-Homs, R. Gonz\'alez, R. Ravelo}

\affiliation{Department of Physics, University of Texas at El
Paso, El Paso, Texas 79968, U.S.A.}

\date{\today}

\begin{abstract}
This study uses classical molecular dynamics to simulate infinite
nuclear matter and study the effect of isospin asymmetry on bulk
properties such as energy per nucleon, pressure, saturation
density, compressibility and symmetry energy. The simulations are
performed on systems embedded in periodic boundary conditions with
densities and temperatures in the ranges $\rho=0.02$ to $0.2 \
fm^{-3}$ and T = 1, 2, 3, 4 and 5 MeV, and with isospin content of
$x=Z/A=0.3$, 0.4 and 0.5. The results indicate that symmetric and
asymmetric matter are self-bound at some temperatures and exhibit
phase transitions from a liquid phase to a liquid-gas mixture. The
main effect of isospin asymmetry is found to be a reduction of the
equilibrium densities, a softening of the compressibility and a
disappearance of the liquid-gas phase transition. A procedure
leading to the evaluation of the symmetry energy and its variation
with the temperature was devised, implemented and compared to mean
field theory results.
\end{abstract}

\maketitle

\section{Introduction}\label{intro}

Investigations of neutron-rich nuclei have recently attracted
attention due to the advent of radioactive beam
facilities~\cite{HIRFL,RIKEN,GSI}.  The goal of such experimental
studies is to extend what is known about the nuclear forces away
from the valley of stability and in the direction of large isospin
asymmetry; this is particularly important to understand the
stability of radioactive nuclei and their reactions, as well as
bulk properties of nuclear matter of relevance in
astrophysics~\cite{Li98,Li01,Dan02,Latt,Stein}.

The study of the role of isospin asymmetry on nuclear forces
predates current efforts.  The fact that light nuclei tend to have
equal numbers of protons and neutrons motivated Bethe and
Weizs\"acker to add an {\it ad hoc} term to their parametrization
of the nuclear binding energy to favor the condition A=2Z, (other
researchers later imposed it to the volume and surface
terms~\cite{myers}); such condition is attributable to the Pauli
exclusion principle which lowers the overall energy of the nucleus
by filling both protons and neutrons to equal levels instead of
different ones.  The isospin symmetric rule, however, is not
satisfied in heavier nuclei (A $\gtrsim$ 40) which tend to have
more neutrons than protons due to a competition between the
short-ranged nuclear force and the Coulomb force: in large nuclei
the distance between nucleons on opposite ends of the nucleus
exceeds the short-range of the attractive p-n force and thus are
not be able to overcome the repulsive p-p Coulomb force unless
more neutrons are added to restore stability.

This interplay between quantum and classical effects illustrates
the difficulty of understanding the role of isospin symmetry in
the nuclear binding energy in terms of first principles.  The
situation becomes more entangled in heavy-ion reactions where the
nucleon energy varies with density and temperature in addition to
isospin.  To incorporate all of these degrees of freedom it is
necessary to replace the Bethe-Weizs\"acker parametrization by a
full-fledged equation of state (EOS).

An intermediate compromise was based on an extension of the liquid
drop formula into non-symmetric isospin values through an additive
term~\cite{tsa09},
$E(\rho,\alpha)=E(\rho,\alpha=0)+E_{Sym}(\rho)\alpha^2
+O(\alpha^4)$ with $\alpha=(N-Z)/A$. Operationally such expression
can be taken as a Taylor expansion of $E(\rho,\alpha)$ in terms of
$\alpha$ about the isospin symmetric point $\alpha=0$ with the
odd-terms in $\alpha$ excluded due to the exchange symmetry
between protons and neutrons of the strong force.  Under this
scheme, the symmetry term is readily obtained through
$E_{Sym}(\rho)=(1/2!)(\partial^2 E/\partial \alpha^2)$ and can be
analyzed as a function of the density and compared to experimental
data (see e.g.~\cite{tsa09,traut,nato}).  In spite of these
efforts, at present the isospin dependence of the nuclear equation
of state is far from being determined; this is particularly true
for the temperature dependence of $E_{Sym}$ which has been less
investigated than its zero temperature counterpart.

A more complete approach is, of course, to use microscopic
theories to develop a complete equation of state with the density,
temperature and isospin degrees of freedom built in from the
start. Theories such as relativistic~\cite{16, 17, 18, 19, 20, 21,
22} and nonrelativistic~\cite{23, 24} Hartree Fock approximations,
as well as relativistic~\cite{baran} and non-relativistic
mean-field models\cite{25,26, 27,28,29,30,31} have been used for
this purpose at T=0 with varying degrees of success; the reader is
directed to~\cite{li,chen} for a comprehensive review of the use
of these techniques in the study of the symmetry energy term.  In
a nutshell, the knowledge we have about the properties of
asymmetric nuclear matter is as good as the techniques used for
solving the nuclear many-body problem, which are far from perfect.

A way around these technical difficulties is through the use of
numerical methods like Monte Carlo, molecular dynamics, lattice
calculations, etc. which are able to construct systems from which
the wanted properties of nuclear matter can be obtained
phenomenologically.  Transport-theory models that have been used
in nuclear reactions can be divided into classical, semiclassical
or quantum. Synoptically, semiclassical models (BUU, IBUU, etc.,
see e.g.~\cite{famiano}) track the time evolution of the Wigner
function under a mean potential to obtain a description of the
probability of finding a particle at a point in phase space. On
the quantum side, the molecular dynamics models (QMD, AMD, IQMD,
etc., see e.g.~\cite{ono,singh,kumar}) solve the equations of
motion of nucleon wavepackets moving within mean fields (derived
from Skyrme potential energy density functional) with a Pauli-like
blocking mechanism imposed and using isospin-dependent
nucleon-nucleon cross sections, momentum dependence interactions,
etc. Although these methods succeed in presenting a reasonable
evolution of density fluctuations in heavy-ion reactions, they
both fail to produce clusters of appropriate quality or
quantity~\cite{famiano}, and their use of hidden adjustable
parameters (width of wavepackets, number of test particles,
modifications of mean fields, effective masses and cross sections,
etc.) makes their findings questionable at best.  As before,
practitioners of this field conclude that more detailed analysis
for the equation of state is needed for the nuclear and the
astrophysical community\cite{kumar13}.

\begin{figure}  
\begin{center}
\includegraphics[width=3.4in]{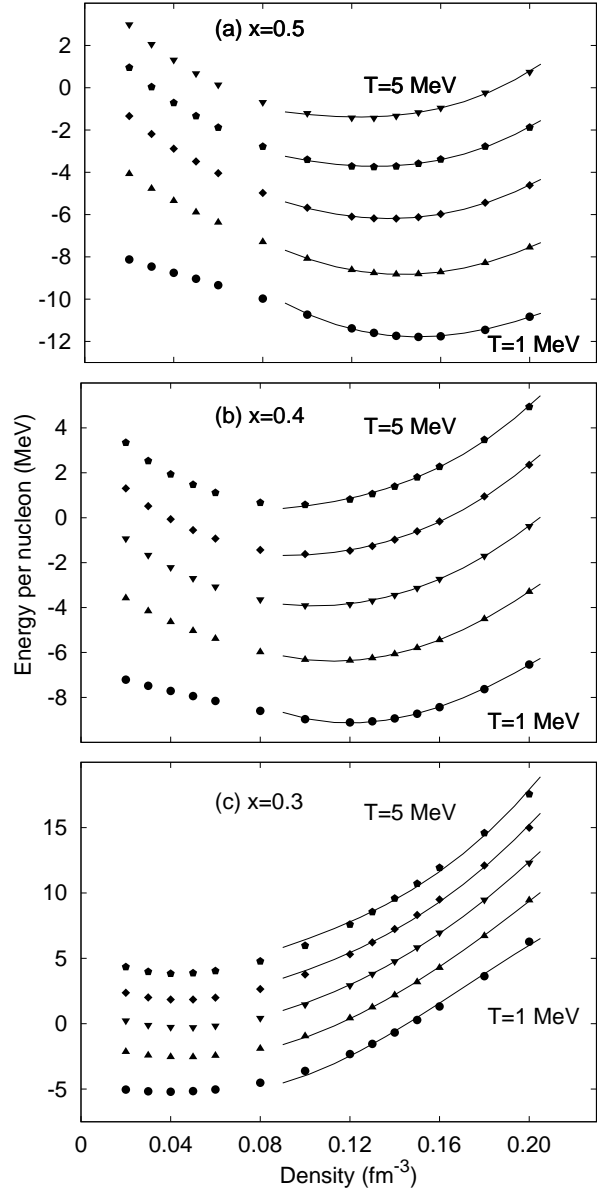}
\end{center}
\caption{Energy per nucleon as a function of the density for three
different isospin contents. In each case the curves correspond to
temperatures ranging from T = 1 MeV (lower curve) to 5 MeV (upper
curve) with the intermediate curves corresponding to 2, 3 and 4
MeV. The lines indicate the fits used in Section~\ref{nse} to
estimate the symmetry energy.} \label{e-d}
\end{figure}

On the other hand, classical molecular dynamics (CMD) models have
been used to study nuclear reactions since decades
ago~\cite{wilets,pandha,lop-lub,dor-ran} and neutron star crusts
more
recently~\cite{horo_lambda,P14,P15,P2012,dor12,dor12A,dor12-2}.
Generally speaking nucleons are treated as classical particles
interacting through pair potentials with their equations of motion
solved numerically with any of the several methods available,
without any adjustable parameters and including all particle
correlations at all levels, i.e. 2-body, 3-body, etc. Indeed the
method can describe nuclear systems ranging from highly correlated
cold nuclei (such as two approaching heavy ions in their ground
state), to hot and dense nuclear matter (nuclei fused into an
excited blob), to phase transitions (fragment and light particle
production), to hydrodynamics flow (after-breakup expansion) and
secondary decays (nucleon and light particle emission).  The only
apparent disadvantage of the CMD is the lack of quantum effects,
such as the Pauli blocking, which at very low excitation energies
stops the method from describing nuclear structure correctly;
fortunately, in collisions the high energy deposition opens widely
the phase space available for nucleons and renders Pauli blocking
practically obsolete.

\begin{figure}  
\begin{center}
\includegraphics[width=3.4in]{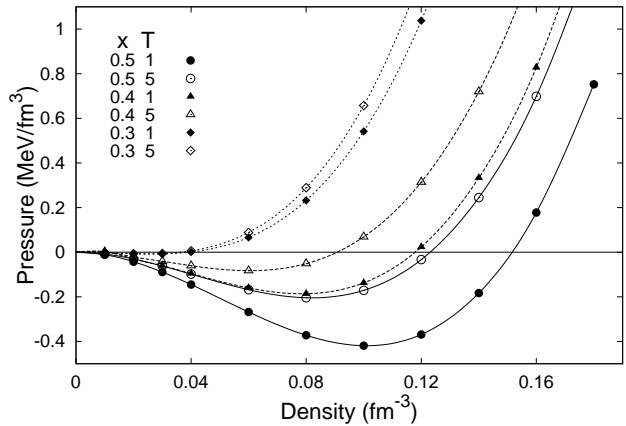}
\end{center}
\caption{Values of the pressure obtained in the molecular dynamics
simulations presented as a function of the density for three
different isospin contents and T = 1 and 5 MeV.} \label{pressure}
\end{figure}

Thus the motivation of this study, to use CMD as a computational
many-body technique to simulate infinite nuclear systems with
varying density, temperature and isospin content to extricate the
isospin dependence of as many nuclear characteristics as possible.
In the following section the model used to obtain the energy,
pressure, saturation density, compressibility and symmetry energy
of infinite nuclear systems at different values of isospin,
density and temperature will be presented. An overview of the
resulting bulk properties of the systems studied is presented in
Section~\ref{bulk}, followed by a discussion of the existence of
phases in asymmetric matter in Section~\ref{phases}, an estimation
of the nuclear symmetry energy in Section~\ref{nse}, a discussion
of the limit of applicability of the classical model in
section~\ref{quantum} and a summary of findings in
Section~\ref{concluding}.

\section{Classical molecular dynamics}\label{cmd}

In this work we use a CMD model with the Pandharipande potentials
which were designed by the Urbana group to reproduce experimental
cross sections in nucleon-nucleon collisions of up to 600
MeV~\cite{pandha}, to mimic infinite systems with realistic
binding energy, density and compressibility and to produce
heavy-ion dynamics comparable to those predicted by the
Vlasov-Nordheim equation.  This parameter-free model has been
successfully used to study nuclear reactions obtaining mass
multiplicities, momenta, excitation energies, secondary decay
yields, critical phenomena and isoscaling behavior that have been
compared to experimental
data~\cite{14a,Che02,16a,Bar07,CritExp-1,CritExp-2,TCalCur,
EntropyCalCur,8a,Dor11}.  More recently, and of interest to the
present work, the model was used to study infinite nuclear systems
at low temperatures~\cite{2013} and in neutron star crust
environments~\cite{dor12,dor12A,dor12-2}.

\begin{figure}
\begin{center}
\includegraphics[width=3.4in]{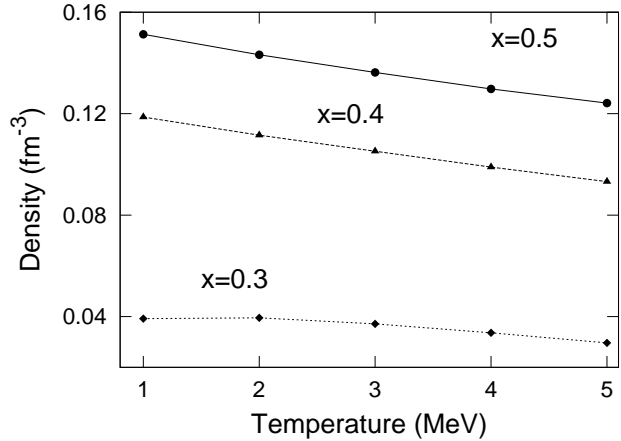}
\end{center}
\caption{Saturation density as a function of the temperature for
three different isospin content.} \label{Min-dens-vs-T}
\end{figure}

The Pandharipande potentials are comprised of an attractive
potential between a neutron and a proton: $V_{np}(r) =
V_{r}exp(-\mu _{r}r)/{r}-V_{r}exp(-\mu_{r}r_{c})/{r_{c}}]$ $-V_{a}
exp(-\mu_{a}r)/{r}-V_{a}exp(-\mu_{a}r_{c})/{r_{c}}$, and a
repulsive interaction between equal nucleons (nn or pp):
$V_{NN}(r)=V_{0}exp(-\mu _{0}r)/{r}-V_{0}exp(-\mu
_{0}r_{c})/{r_{c}}$.  The range of these potentials is limited to
a cutoff radius of $r_c=5.4$ fm after which they are set to zero.
The parameters $V_r$, $V_a$, $V_0$, $\mu_r$, $\mu_a$ and $\mu_0$
were phenomenologically adjusted by Pandharipande to yield a cold
nuclear matter saturation density of $\rho_0=0.16$ fm${}^{-3}$, a
binding energy $E(\rho_0)=-16$ MeV/nucleon and a compressibility
of about 250 MeV for their ``medium'' model, which is the one used
here.

At a difference from a previous study of nuclear matter at low
temperatures~\cite{2013} where this model was used to obtain a
description of the equilibrium structures (i.e. the ``pasta''), in
the present case we are interested on creating systems with
different values of isospin content that will allow us to extract
the isospin dependence of physical observables such as the energy
per nucleon, pressure, equilibrium density, compressibility and
the symmetry energy. With this in mind, and with an eye on future
studies of transport coefficients, the molecular dynamics code
used is based on the Nos\'e-Hoover equations of motion which add
to the newtonian mechanics the effect of a heat
reservoir~\cite{ravelo95}.

The addition of the heat flow variable to the classical equations
of motion results in the Nos\'e-Hoover equations of motion which
can be integrated by St{\o}rmer finite differences.  In principle
this approach corresponds to a canonical ensemble and does not
conserve energy which is added or removed by the heat reservoir;
configurations in thermal equilibrium, however, can be achieved
faster than with the usual microcanonical formalism of newtonian
mechanics and an Andersen's thermostat~\cite{andersen}.

\begin{figure}  
\begin{center}
\includegraphics[width=3.4in]{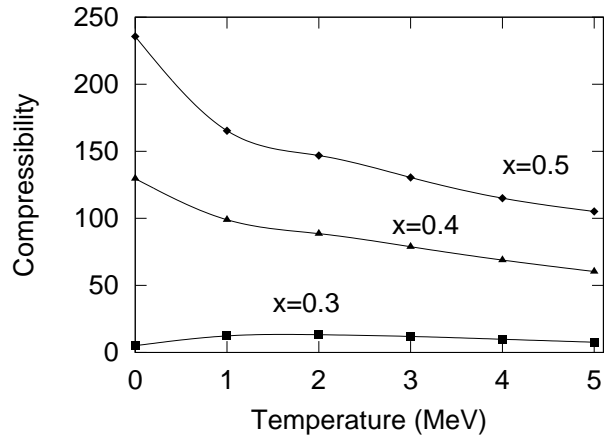}
\end{center}
\caption{Compressibility as a function of the temperature for three different isospin content.} \label{CompVsTemp}
\end{figure}

To mimic an infinite system $A=2000$ nucleons were placed in cubic
cells under periodic boundary conditions. We focus on systems with
isospin content of $x=Z/A=0.3$, 0.4 and 0.5, where $Z$ is the
number of protons. The number densities were enforced by placing
the nucleons in cubical boxes with sizes selected to adjust the
density. The temperatures of the systems studied are T = 1, 2, 3,
4, and 5 MeV, and their densities were selected to be around and
below the corresponding saturation densities values, which vary
with isospin content and temperature. The procedure followed is
straightforward: the nucleons are placed at random within the cell
avoiding overlaps (i.e. interparticle distances smaller than 0.1
fm) and endowed with a Maxwell-Boltzmann velocity distribution
corresponding to the desired temperature.  The system then is
rapidly evolved until the temperature is maintained within 1$\%$.
After reaching thermal equilibrium, the system continues evolving
and its information at selected time steps (nucleon positions and
momenta, energy per nucleon, pressure, temperature, density, etc.)
is stored for future analysis. The calculations were carried out
in the High Performance Computing Center of the University of
Texas at El Paso which has a beowulf class of linux clusters with
285 processors.

\section{Bulk properties of nuclear matter}\label{bulk}

Following the procedure outlined before, the energy per nucleon
(kinetic plus potential) was calculated for systems with $x=0.3$,
0.4 and 0.5. Figure~\ref{e-d} shows the variation of the energy as
a function of the density at temperatures T = 1, 2, 3, 4 and 5
MeV; each point represents the average of 200 thermodynamically
independent configurations, the average of the standard deviations
is 0.036 MeV which is smaller than the points used. In all cases
the curves show the characteristic ``$\cup$'' shapes around their
corresponding saturation density (minimum of the $\cup$). It is
easy to see that bound matter (i.e. with negative binding energy)
exists in all three isospin cases although at high temperatures
the systems become unbound. Also seen in the plots is a departure
from the $\cup$ shapes at low densities signaling a transitions
from a uniform continuous medium around saturation densities  to
non-homogeneous media at subsaturation densities; this effect is
clearly noticeable for $x=0.5$ and 0.4 but not for 0.3, point
which will be discussed further in the following section.

\begin{figure}  
\begin{center}
\includegraphics[width=3.4in]{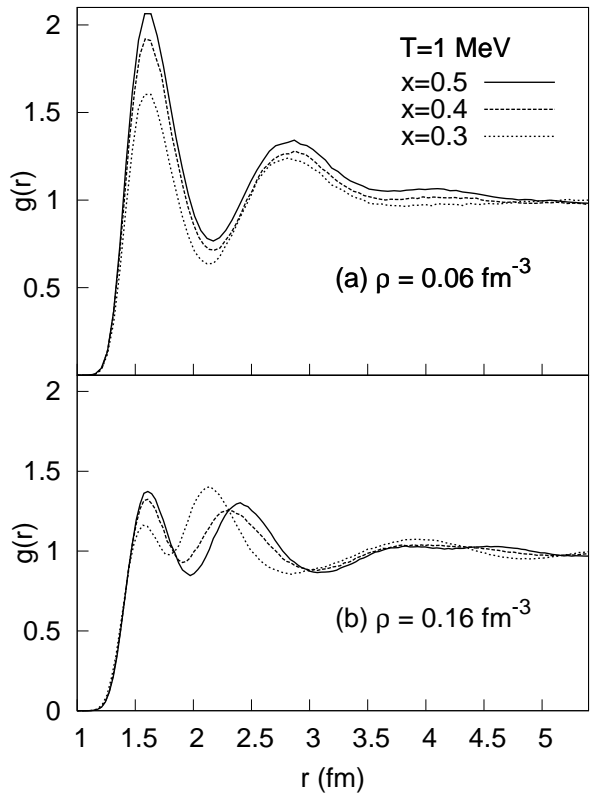}
\end{center}
\caption{Radial distribution functions of systems with $x=0.3$,
0.4 and 0.5 at T = 1 MeV and densities $\rho=0.06 \ fm^{-3}$ in
panel (a) and $\rho=0.16 \ fm^{-3}$ in panel (b).} \label{rdf}
\end{figure}

The CMD calculations also yielded the pressure of the systems at
each values of $(T,\rho,x)$. Figure~\ref{pressure} shows the
pressure versus density curves for T=1 and 5 MeV for the three
isospin contents $x=0.3$, 0.4 and 0.5. As seen in Figs.~\ref{e-d}
and~\ref{pressure}, the equilibrium densities correspond to the
minima of the energy-density curves as well as to the zero
pressure points; we estimate the values of the saturation density
through a least-squares fit of a quadratic polynomial of
$E(T,\rho)$ around the minimum of each curve;
Figure~\ref{Min-dens-vs-T} shows the variation of the saturation
density with the temperature for the three values of $x$. As
expected, the trend for symmetric matter tends to $\rho_0 = 0.16 \
fm^{-3}$ as T goes to zero as expected for infinite cold nuclear
matter. The case of $x=0.3$ appears to have a very low saturation
density with little temperature variation.

The compressibility at saturation density can also be obtained
from the previous fit through $K(T,\rho)=9\rho^2\left[
\partial^2 E/\partial\rho^2 \right]_{\rho_0}$.
Figure~\ref{CompVsTemp} shows the values obtained for $K(T,\rho)$
for the cases studied; as a reference the values at T = 0 obtained
in a previous work~\cite{2013} are also included. As it can be
seen, the compressibility decreases drastically with $x$, and it
is further reduced (by about 30\%) as T increases from 1 to 5 MeV
for the systems with $x=0.4$ and 0.5, while it remains extremely
soft in the neutron-rich system of $x=0.3$.

\begin{figure}
\begin{center}
\includegraphics[width=3.4in]{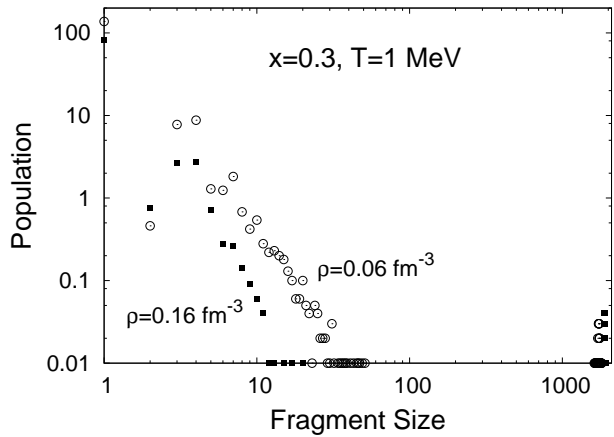}
\end{center}
\caption{Mass distribution obtained with $x=0.3$, T = 1 MeV at
$\rho=0.06$ and $0.16 \ fm^{-3}$.} \label{MassDist}
\end{figure}

\section{Phases in asymmetric matter}\label{phases}

Figure~\ref{e-d} shows the transition from a continuous phase to
an amorphous one, this phenomenon was first pointed out for cold
symmetric matter in the Thomas-Fermi calculation of Williams and
Koonin~\cite{koonin}, and was recently explored at non-zero
temperatures (T$<$1 MeV) in a CMD microcanonical study~\cite{2013}
using this and two other potentials.  In the case of T = 1 MeV
(cf. Fig.~\ref{e-d}) it is easy to see the transition from a
smooth $\cup$ shape to an extraneous curve at $\rho \approx 0.10 \
fm^{-3}$ for $x=0.5$, and at $\rho \approx 0.08 \ fm^{-3}$ for
$x=0.4$; the behavior persists in these systems at higher
temperatures but appears to be absent for the $x=0.3$ case.  As
discussed in detail for the symmetric case in~\cite{2013}, the
smooth $\cup$ shape corresponds to a uniform phase (crystal-like
at low temperatures and liquid-like at higher temperatures) and
the lower-density separating part signals the existence of a
non-homogeneous structure (such as a ``pasta'' at low temperatures
and a liquid-gas mixture at higher temperatures).

\begin{figure}  
\begin{center}
\includegraphics[width=3.4in]{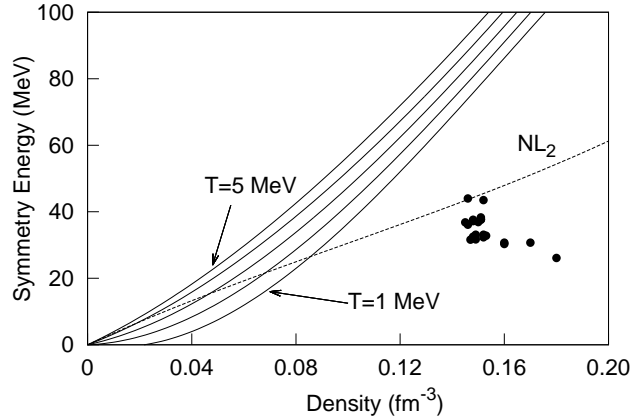}
\end{center}
\caption{The solid lines show the symmetry energies obtained from
the CMD values of $E(T,\rho,\alpha)$ for T = 1, 2, 3, 4 and 5 MeV
as explained in the text. For comparison, the dashed line ($NL_2$)
and the points all show values of the symmetry energy obtained
with different relativistic Hartree calculations.}
\label{ESym-Dens}
\end{figure}

\begin{figure}  
\begin{center}
\includegraphics[width=3.4in]{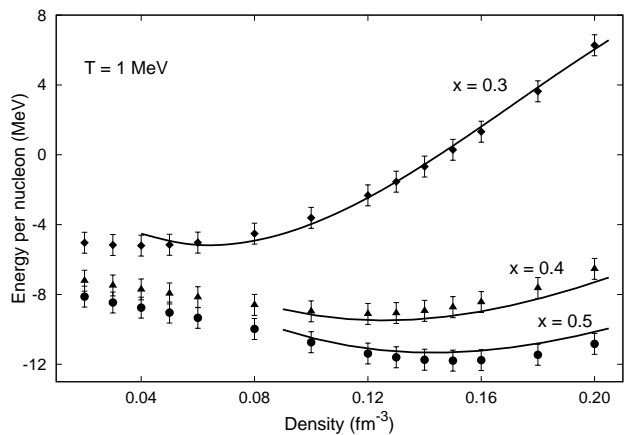}
\end{center}
\caption{The lines show the fits of $E(T,\rho,\alpha)$ for T=1 MeV
that lead to a symmetry energy identical to $NL_2$ in
Fig,~\ref{ESym-Dens}.} \label{Fig-Fit-T-1.eps}
\end{figure}

The different behavior of the systems with $x=0.3$ would suggest
that at low temperatures the system would never enter a liquid-gas
mixture region and would always stay in a liquid-like continuous
medium down to very low densities. To investigate this we examined
the structures formed at T = 1 MeV through the radial distribution
function (RDF) and their mass distribution.

The radial distribution function, $g(r)=\rho(r)/\rho$, (cf.
Figure~\ref{rdf}) was obtained from averaging 200 systems at T=1
MeV at a liquid density ($\rho=0.16 \ fm^{-3}$, bottom panel) and
at a liquid-gas mixture density ($\rho=0.06 \ fm^{-3}$, top
panel).  The strengths of the nearest-neighbor peaks show that at
low densities nucleons tend to be more correlated than at higher
densities indicating that at $\rho=0.06 \ fm^{-3}$ the main
contribution at short distances is from nucleons in droplets,
while at $\rho=0.16 \ fm^{-3}$ the larger nucleon mobility reduces
such correlation. This is also observed in the second-neighbor
peaks which appear at the same distance for all values of $x$ at
$\rho=0.06 \ fm^{-3}$ but not at $\rho=0.16 \ fm^{-3}$ indicating
again a reduced mobility of nucleons in the droplets.  The growth
of the second-neighbors peak over the first one for $x=0.3$ is due
to the large number of nn repulsive interactions which exceed the
smaller number of np attractive interactions.

It is instructive to explore the mass distributions attained by
the systems in these two density regimes.  In all cases the mass
distribution at $\rho=0.06 \  fm^{-3}$ showed an over-abundance of
intermediate mass fragments over those obtained at $\rho=0.16 \
fm^{-3}$. Figure~\ref{MassDist} shows that for $x=0.3$ at T = 1
MeV the system contains more intermediate-mass droplets at
$\rho=0.06 \ fm^{-3}$  than at $0.16 \ fm^{-3}$ where more matter
remained in a continuous medium; this figure was obtained by
averaging 200 systems at similar conditions and the clusters were
identified through a minimum spanning tree method.

The results presented in this section thus indicate that the
transition from an homogeneous medium at near-saturation densities
to a non-homogeneous one at lower densities continues to exist in
asymmetric nuclear matter with $x=0.4$ and 0.3 at low temperatures
($T\lesssim 2 MeV$); in other words, phases appear to be are alive
and well in these cases of high isospin asymmetry.  A related
observation is that these systems become unbound at temperatures
as low as T = 2 MeV for $x=0.3$; observation that is in line with
previous findings of the unboundness of pure neutron matter within
the nuclear Thomas-Fermi model~\cite{myers95}, and it is probably
connected to the shift of the saturation density to exceedingly
low values for low $x$.

\section{Nuclear symmetry energy}\label{nse}

The previous scheme can be used to extract the nuclear symmetry
energy through $E_{Sym}(\rho)=(1/2!)\left[\partial^2 E/\partial
\alpha^2\right]_{\alpha=0}$. This can be done by inserting the
$\alpha$ dependence into an expression of the type
$E(T,\rho,\alpha)=E_1(T,\alpha)\rho+E_2(T,\alpha)\rho^2+E_3(T,\alpha)\rho^3$
and adjusting the parameters $E_1(T,\alpha)$, $E_2(T,\alpha)$ and
$E_3(T,\alpha)$ to fit the values of $E(T,\rho,x)$ for each T and
$x$ in the $\cup$ region, namely from $\rho=0.09$ to $0.2 \
fm^{-3}$ for $x=0.4$ and 0.5, and $\rho=0.04$ to $0.2 \ fm^{-3}$
for $x=0.3$.  The resulting fits are shown as solid lines in
Figure~\ref{e-d}.

More explicitly, using the data for $x=0.5$, 0.4 and 0.3 and
remembering that $\alpha=(N-Z)/A=1-2x$, it is possible to obtain,
for instance, values for $E_1(T,\alpha=0)$, $E_1(T,\alpha=0.2)$
and $E_1(T,\alpha=0.4)$ and approximate
$E_1(T,\alpha)=E_{10}(T)+E_{12}(T)\alpha^2+E_{14}(T)\alpha^4$,
where the coefficients $E_{10}(T), \ E_{12}(T)$ and $E_{14}(T)$
can be obtained by solving the resulting system of coupled
equations. Using similar expressions for $E_2(T,\alpha)$ and
$E_3(T,\alpha)$ it is simple to obtain
$E_{Sym}(T,\rho)=E_{12}(T)+E_{22}(T)\rho+E_{32}(T)\rho^2$. The
resulting $E_{Sym}$ are presented in Figure~\ref{ESym-Dens} as a
function of the density for T=1, 2, 3, 4 and 5 MeV.

The smooth dependence of $E_{Sym}$ with the density is reminiscent
of previous results obtained with microscopic field theories. For
comparison we show the symmetry energy obtained by Chen at
al.~\cite{chen} using a relativistic Hartree calculation (curve
labeled ``$NL_2$'') along with values obtained with other field
theories all plotted at their corresponding saturation densities
(see~\cite{chen} for complete details).

Another result worth mentioning is the smooth temperature
dependence of $E_{sym}(T,\rho)$ which, interestingly, has a trend
opposite to previous findings. In the past the temperature
dependence of $E_{sym}$ has been studied using an equation of
state obtained through a virial expansion at T = 2, 4 and 8
MeV~\cite{horo-s}, and through a self-consistent model using
various effective interactions at temperatures ranging up to 50
MeV~\cite{xu}; in both of these calculations an inverse
temperature dependence was found, that is $E_{sym}$ decreased as T
increased.  From a different perspective, the CMD results appear
to be in line with those obtained by de Lima and
Randrup~\cite{randrup} with a modified Seyler-Blanchard model,
which show an overall increase of the free energy through changes
of the volume and surface coefficients and nuclear surface tension
in the range of $T\lesssim5$ MeV.

It stands to reason that the differences between the CMD results
and those of field theories reflect the distinct underlying
assumptions of the models. A major difference is that the field
theories were adapted to reproduce the binding energies and
charged radii of stable (i.e. nearly isosymmetric) finite nuclei
in their ground-state while the CMD results stem out of a
generalized analysis of heated infinite systems at varying isospin
content. Another significant difference is that the $E_{sym}$
extracted from the CMD results is not based on the extraneous
partition of the liquid drop formula into an isospin symmetric
part and an isospin asymmetric additive term; indeed the CMD
results stem from the holistic effects the density, temperature
and isospin dependence have on the energy simultaneously. Other
differences are the amount of interactions that each method
incorporates, while field theories are usually limited to the
ladder diagrams, the CMD contains all many-body interactions.
Fortunately, as it will be discussed in more detail in the
section~\ref{quantum}, the lack of quantum effects of CMD appears
not to play a major role in the validity of these results.

However, independent of all of these differences between the
models, there is another important factor. Unfortunately --or
perhaps, fortunately--, the procedure used to obtain $E_{sym}$
from the CMD results involves fitting algorithms that contain an
intrinsic variability (due to the specific points included in the
fit, error bars, etc.) that yields a margin to play with the fit
of $E(T,\rho,\alpha)$ which in turn produces various functional
forms of $E_{sym}(T,\rho)$. In fact, it is possible to work
``backwards'' (using multivariate regression analysis) and find
relatively close fits of $E(T,\rho,\alpha)$ that produce an
specific form of $E_{sym}(T,\rho)$. Figure~\ref{Fig-Fit-T-1.eps},
for instance, shows fits of $E(T,\rho,\alpha)$ for T = 1 MeV that
fall within 0.7 MeV of the CMD data points (indicated by the data
bars) and yield exactly the $NL_2$ symmetry energy of
Figure~\ref{ESym-Dens}. Viewing this on the positive side, these
results signal a possible compatibility between the CMD results
and other theories. But viewing it on the negative side, this
variability effectively rules out the procedure used for an exact
extraction of the nuclear symmetry energy and limits its scope to
more general ``ball park'' estimates.

\section{Quantum caveats}\label{quantum}
As explained in the Introduction, classical molecular dynamics
lacks all quantum effects, such as the Pauli blocking, which at
low excitation energies could produce an incorrect energy
distribution. The question then arises, how does this deficiency
of CMD affect the results of this investigation? In the
Introduction it was stated that for high excitation energies the
phase space available for nuclei would be so ample that it would
render Pauli blocking practically obsolete. In this section such
statement is quantified in a bit more detail.

\begin{figure}  
\begin{center}
\includegraphics[width=3.4in]{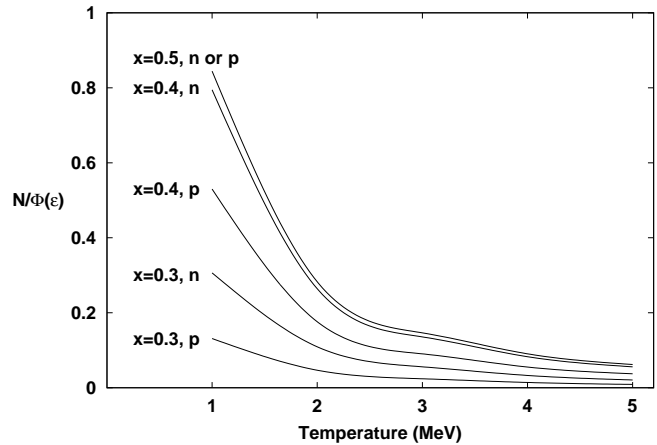}
\end{center}
\caption{Values of $N/\Phi(\epsilon)$ as a function of temperature
for $x=0.3$, 0.4 and 0.5 and for protons (p) and neutrons (n).}
\label{level}
\end{figure}

Quantum effects affect the behavior of many body systems on, at
least, two fronts: energy distribution and wave mechanics.  From
the point of view of the energy, in bound clusters the energy of
individual nucleons becomes discrete and the distribution of
energy levels is ruled by Fermi-Dirac statistics with Pauli
exclusion principle further regulating the occupation of such
levels. As stated before, at high excitation energies the density
of states becomes so finely dense that Pauli blocking is rendered
obsolete, such limit can be said to exist whenever the number of
quantum states available to a nucleon at a given temperature is
much greater than the number of nucleons. According to statistical
mechanics this previous condition is $\Phi(\epsilon) \gg N$ and,
for the simple case of a particle in a box where
$\Phi(\epsilon)={{\pi V}\over{6}} \left({{8 M \epsilon} \over
{h^2}}\right)^{3/2}$ and $\epsilon =3T/2$, it can be approximated
by~\cite{mac}
\begin{equation}
N/\Phi(\epsilon)=\sqrt{{\pi}\over{6}} {{\rho h^3}\over{(2 \pi M
T)^{3/2}}} \ll 1
 \ , \label{nlam1}
\end{equation}
where M is the nucleon mass and $\rho$ is the neutron or proton
number densities at saturation.  Using the values of $\rho$ from
Figure~\ref{Min-dens-vs-T} along with $\rho_n=(1-x)\rho$ and
$\rho_p=x\rho$ for neutrons and protons respectively,
$N/\Phi(\epsilon)$ yields the values shown in Figure~\ref{level},
which demonstrate that condition (\ref{nlam1}) is fully satisfied
in all of the cases considered in this study except, perhaps, for
$x\ge 0.4$ at T $\approx$ 1 MeV where $N \gtrsim \Phi(\epsilon)$
and the condition is marginally satisfied.

\begin{figure}  
\begin{center}
\includegraphics[width=3.4in]{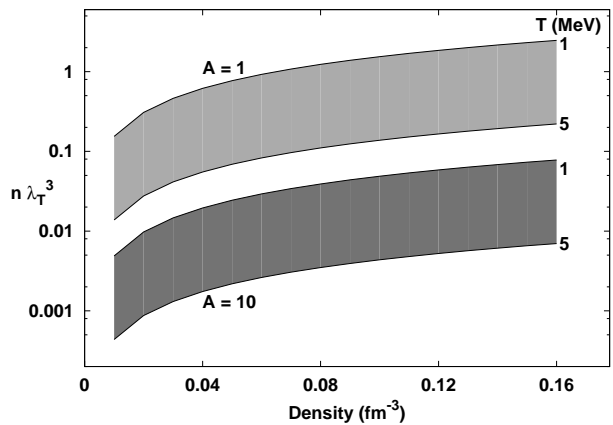}
\end{center}
\caption{Values of $\rho \lambda_T^3$ as a function of density and
for a range of temperatures and for two values of the masses.}
\label{Fignlam}
\end{figure}

A second test of the validity of the classical approach has to do
with the wave features of the particles.  It is known that wave
mechanics yields to classical mechanics whenever the mean
inter-particle distance is much larger than the mean thermal de
Broglie wavelength $\lambda_T$. Using $(V/N)^{1/3}=\rho^{-1/3}$ as
the inter particle distance, this  condition yields to the
inequality~\cite{kittel}
\begin{equation}
\rho \lambda_T^3 = {{\rho h^3}\over{(2 \pi M T)^{3/2}}} \ll 1
 \ . \label{nlam}
\end{equation}
We notice that condition (\ref{nlam}) is practically identical to
condition (\ref{nlam1}) except for a small numerical factor of
order 1. In this case, however, M stands for the nuclei masses and
$\rho$ for the overall system's density. Figure~\ref{Fignlam}
shows the values of $\rho \lambda_T^3$ as a function of density
and for masses A = 1 (top band) and 10 (lower band). The shaded
regions indicate the values obtained for each of the two masses
for the range of temperatures 1 MeV $\le$ T $\le$ 5 MeV; higher
masses produce bands with values of $\rho \lambda_T^3$ that are
even smaller. It is easy to see that condition (\ref{nlam}) is
fully satisfied by all size particles at all of the densities and
temperatures considered in the present study except, perhaps, for
free nucleons at T $\approx$ 1 MeV and $\rho > 0.1 \ fm^{-3}$
where $\rho \lambda_T^3 \approx 1$ and the condition is marginally
satisfied.

Figures~\ref{level} and~\ref{Fignlam} thus provide a degree of
confidence over the CMD results. Indeed Fig.~\ref{level} provides
certainty over the energy distribution attained by nucleons in
nuclei near saturation density, while Fig.~\ref{Fignlam} confirms
the validity of the dynamics of the system (and hence its
thermodynamics) in the liquid-gas phase. Other quantum effects,
such as collective excitations, superfluidity and
superconductivity, etc. are known to occur at lower temperatures,
near the ground state of the nucleus, and are unlikely to affect
the results of the present calculation.

\section{Concluding remarks}\label{concluding}

In this study we used CMD to simulate infinite nuclear systems
with varying density, temperature and isospin content to extract
the isospin dependence of the energy per nucleon, pressure,
saturation density, compressibility and symmetry energy. We
studied systems with 2000 nucleons embedded in periodic boundary
conditions with densities and temperatures in the ranges
$\rho=0.02$ to $0.2 \ fm^{-3}$ and T = 1, 2, 3, 4 and 5 MeV, and
with isospin content of $x=Z/A=0.3$, 0.4 and 0.5.

The results obtained for the energy per nucleon (cf.
Fig.~\ref{e-d}) indicate that symmetric ($x=0.5$) and asymmetric
matter ($x=0.3$ and 0.4) can be self bound for certain values of T
and $\rho$. The equilibrium densities obtained from the minima of
the energy-density curves indicate that as T increased from 1 to 5
MeV, the saturation densities varied from $\rho_0=0.16 \ fm^{-3}$
to $0.12 \ fm^{-3}$ for isospin symmetric matter, and from
$\rho_0\approx 0.12 \ fm^{-3}$ to $0.09 \ fm^{-3}$ for matter with
$x=0.4$.

The compressibility around saturation density was determined from
the $E(T,\rho)$ data and it was found that isospin asymmetry
softens nuclear matter by a factor of about $50\%$ as $x$ drops
from 0.5 to 0.4 or from 0.4 to 0.3. It was also observed that the
compressibility is further reduced (by about 30\%) as T increases
from 1 to 5 MeV for the systems with $x=0.4$ and 0.5.

The existence of phases was identified by examining the density
dependence of the energy per nucleon. At all isospin asymmetries
the energy $E(T,\rho)$ showed $\cup$ shapes around the saturation
densities characteristic of a continuous liquid-like phase. At
sub-saturation densities the energy-density curves of $x=0.4$ and
0.5 departed from the uniform liquid phase signaling a transition
to a non-homogeneous one, presumably a liquid-gas mixture.

The larger asymmetric case of $x=0.3$ deserves a special mention.
Such systems become unbound at all densities for temperatures as
low as T = 3 MeV, and exhibited very low saturation densities and
extremely small compressibility at all temperatures.  These
assertions were ratified by the measurements of the the
zero-pressure densities (cf. Figure~\ref{pressure}). Although
these findings would suggest that at low temperatures the system
would never enter a liquid-gas mixture region and would always
stay in a liquid-like continuous medium down to very low
densities, a close examination of the structures formed at T = 1
MeV through the radial function and mass distribution at
$\rho=0.06$ and $\rho=0.16 \ fm^{-3}$ indicate that phases are
alive and well in this highly isospin symmetric case.

Turning now to the nuclear symmetry energy, a procedure to obtain
$E_{Sym}(T,\rho)$ from the CMD values of $E(T,\rho,\alpha)$ was
devised and implemented. The results show both a smooth dependence
of the symmetry energy with the density and the temperature.
Unfortunately, the statistical variations of the energy density
are not small enough as to pinpoint the symmetry energy with
satisfactory accuracy.

Comparing to previous studies, our dynamical results confirm
certain previous predictions while extending them to higher
temperatures or other values of isospin asymmetry. For instance,
the shapes of $E(T,\rho,\alpha)$ resemble closely the predictions
of Skyrme–-Hartree–-Fock and relativistic mean-field calculations
for zero temperature~\cite{Tanihata}.  Likewise, the softening of
$K$ with excitation energy appears to be consistent with IQMD
calculations of intermediate-mass fragments multiplicities in
simulations of $^{197}Au+^{197}Au$ at 600 MeV/A~\cite{kumar13} and
with BUU calculations of the traverse component of the elliptic
flow of similar reactions at 1 GeV/A~\cite{Dan02}, both
observables presumably correspond to higher excitations. As
mentioned before, the low binding energy of $x=0.3$ matter is in
line with the lack of binding of neutron-rich~\cite{Tanihata} and
pure neutron matter~\cite{myers95} at zero temperature.

Along the same lines, our results for the symmetry energy are
reminiscent of relativistic and non-relativistic Hartree
calculations~\cite{chen} and, in fact, can be made to fully agree
with such calculations.  Characteristics such as the temperature
dependence of $E_{sym}(T,\rho)$ appear to be in agreement with the
variations of liquid-drop terms and nuclear surface tension in the
range of $T\lesssim5$ MeV~\cite{randrup} but in disagreement with
trends obtained by other theories~\cite{horo-s,xu}.

In summary, the CMD model indeed helps to understand the role of
isospin on several nuclear properties. Besides corroborating
previous studies, this work extends some of their results to other
values of isospin content and non-zero temperatures. Findings that
we believe are new are the temperature variation of the saturation
density and compressibility for isospin asymmetric matter, certain
details of the existing phases at $x=0.3$ and 0.4, as well as a
new procedure to estimate $E_{sym}(T,\rho)$ from kinetic
simulations.

In the future this study will be extended to lower temperatures
(where crystalline structures exist) and other transport
properties such as speed of sound, diffusion coefficients, and to
finite systems.

\begin{acknowledgments}

This study was financed by the National Science Foundation grant
NSF-PHY 1066031 and by DOE's Visiting Faculty Program. The authors
acknowledge helpful conversations with J{\o}rgen Randrup and thank
the warm hospitality of the Nuclear Theory Group of the Nuclear
Science Division of The Lawrence Berkeley National Laboratory
where this work was carried out from beginning to end.
\end{acknowledgments}

\end{document}